\documentclass[%
 reprint,
 amsmath,amssymb,
 aps,
 prb,
 superscriptaddress
]{revtex4-2}
\usepackage[utf8]{inputenc}

\usepackage{graphicx,xcolor}% Include figure files
\usepackage{dcolumn}% Align table columns on decimal point
\usepackage{bm}% bold math
\usepackage[percent]{overpic}
\usepackage{physics}
\usepackage{mathtools}

\definecolor{darkGreen}{RGB}{0,110,0}
\definecolor{darkBlue}{RGB}{0,0,130}

\renewcommand{\pv}{p_{\mathrm{value}}}

\usepackage[colorlinks,citecolor=darkGreen,linkcolor=darkBlue,%
urlcolor=blue,hyperindex]{hyperref}

\renewcommand{\vec}{\boldsymbol}

\usepackage{bm}% bold math

\begin{document}

\title{Wave function network description and Kolmogorov complexity of quantum many-body systems}

\author{T. Mendes-Santos}
\affiliation{Theoretical Physics III, Center for Electronic Correlations and Magnetism,
Institute of Physics, University of Augsburg, 86135 Augsburg, Germany}
\author{M. Schmitt}
\affiliation{Forschungszentrum J\"ulich GmbH, Peter Gr\"unberg Institute, Quantum Control (PGI-8), 52425 J\"ulich, Germany}
\author{A. Angelone}
\affiliation{Sorbonne Université, CNRS, Laboratoire de Physique Théorique de la Matière Condensée, LPTMC, F-75005 Paris, France}

\author{A. Rodriguez}
\affiliation{The Abdus Salam International Centre for Theoretical Physics (ICTP), Strada Costiera 11, 34151 Trieste, Italy}
\affiliation{Dipartimento di Matematica e Geoscienze, Università degli Studi di Trieste, via Alfonso Valerio 12/1, 34127, Trieste, Italy}
\author{P.~Scholl}
\affiliation{Universit\'e Paris-Saclay, Institut d'Optique Graduate School,\\
CNRS, Laboratoire Charles Fabry, 91127 Palaiseau Cedex, France}
\affiliation{California Institute of Technology, Pasadena, CA 91125, USA}
\author{H.~J.~Williams}
\affiliation{Department of Physics, Durham University, South Road, Durham DH1 3LE, UK}
\author{D.~Barredo}
\affiliation{Universit\'e Paris-Saclay, Institut d'Optique Graduate School,\\
CNRS, Laboratoire Charles Fabry, 91127 Palaiseau Cedex, France}
\affiliation{Nanomaterials and Nanotechnology Research Center (CINN-CSIC), Universidad de Oviedo (UO), Principado de Asturias, 33940 El Entrego, Spain}
\author{T. Lahaye}
\affiliation{Universit\'e Paris-Saclay, Institut d'Optique Graduate School,\\
CNRS, Laboratoire Charles Fabry, 91127 Palaiseau Cedex, France}
\author{A. Browaeys}
\affiliation{Universit\'e Paris-Saclay, Institut d'Optique Graduate School,\\
CNRS, Laboratoire Charles Fabry, 91127 Palaiseau Cedex, France}
\author{M. Heyl}
\affiliation{Theoretical Physics III, Center for Electronic Correlations and Magnetism,
Institute of Physics, University of Augsburg, 86135 Augsburg, Germany}
\author{M. Dalmonte}
\affiliation{The Abdus Salam International Centre for Theoretical Physics (ICTP), Strada Costiera 11, 34151 Trieste, Italy}
\affiliation{SISSA — International School of Advanced Studies, via Bonomea 265, 34136 Trieste, Italy}

\begin{abstract}
Programmable quantum devices are now able to probe wave functions at unprecedented levels. 
This is based on the ability to project the many-body state of atom and qubit arrays onto a measurement basis which produces snapshots of the system wave function. Extracting and processing information from such observations remains, however, an open quest. One often resorts to analyzing low-order correlation functions - that is, discarding most of the available information content. 
Here, we introduce wave function networks - a mathematical framework to describe wave function snapshots based on network theory. 
For many-body systems, these networks can become scale free - a mathematical structure that has found tremendous success and applications in a broad set of fields, ranging from biology to epidemics to internet science. 
We demonstrate the potential of applying these techniques to quantum science by introducing protocols to extract the Kolmogorov complexity corresponding to the output of a quantum simulator, and implementing tools for fully scalable cross-platform certification based on similarity tests between networks.
We demonstrate the emergence of scale-free networks analyzing experimental data obtained with a Rydberg quantum simulator manipulating up to 100 atoms. Our approach illustrates how, upon crossing a phase transition, the simulator complexity decreases while correlation length increases - a direct signature of build up of universal behavior in data space. Comparing experiments with numerical simulations, we achieve cross-certification at the wave-function level up to timescales of 4$\mu$s with a confidence level of 90\%, and determine experimental calibration intervals with unprecedented accuracy.
Our framework is generically applicable to the output of quantum computers and simulators with {\it in situ} access to the system wave function, and requires probing accuracy and repetition rates accessible to most currently available platforms.

\end{abstract}

\maketitle

\section{Introduction}

Harnessing and probing many-body systems at the single particle/qubit level are hallmark features of present-day quantum simulators and computers~\cite{gross2017quantum,Browaeys2020,blais2021circuit,monroe2021programmable}. One of the most drastic demonstrations of these tools is the possibility of taking a large number of 'photos' of a many-body system, obtained via projective measurements of the full many-body wave function.  
While this flood of available observations could be seen as a blessing, it immediately encounters practical as well as conceptual challenges: how can this large amount of information be processed, without {\it a priori} discarding some (in the data science language, before performing a dimensional reduction)? What can one learn, that is, e.g., not available by utilizing low-order correlation functions? Answering these questions requires a structured, mathematical understanding of the experimental wave function snapshots, that addresses the {\it information limbo} between traditional many-body theory based on few-points correlation functions~\cite{mahan2013many}, and full-fledged - but experimentally limited to few-particle systems - tomographic methods~\cite{nielsen2002quantum}. 

\begin{figure*}[th]
	\includegraphics[width=1.9\columnwidth]{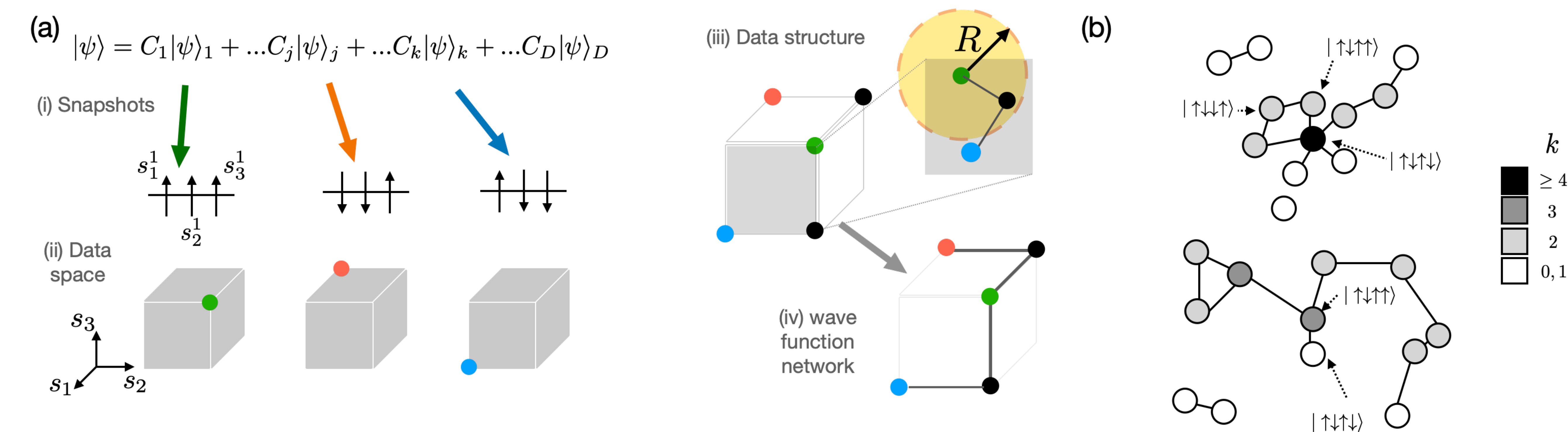}
	\caption{{Network description of many-body wave function snapshots. Panel {\it a)}: construction of the network. First, samples of a wave function are collected (i) and individually mapped onto the target data space (ii). All data are then merged into a single data structure (iii), that defines a set of points in the configuration data space. This data structure is then mapped onto the corresponding wave function network (iv) by drawing links in the network according to a cutoff distance $R$ that is determined by the data structure and the choice of metric (see text). Panel {\it b)} physical interpretation of the network structure.  Within the network, the number of neighbors of given points follows a specific distribution. Points with a large number of links $k$ (i.e., large number of points within $R$) are hubs, and are indicated in darker colors. As an example, taking snapshots of a classical antiferromagnet below its critical temperature will feature the antiferromagnetic state as a hub (top cartoon), while doing so well above its critical temperature will lead to a graph with no hubs and random connections (bottom cartoon).
	}
  }
	\label{fig:Scheme}
\end{figure*}

Here, we develop a theoretical framework to characterize and classify experimentally accessible collections of wave-function snapshots utilizing network theory, that is scalable and allows one to retain all available information.
The backbone of our method is a mapping between collections of wave-function snapshots, and a 'wave-function' network, schematically depicted in Fig.~\ref{fig:Scheme}, that is applicable to spin-, bosonic, and fermionic systems. Utilizing well-established tools in network theory we unravel several key characteristics of the underlying quantum wave function, that are inaccessible by conventional means.

The pivotal finding is that the resulting quantum wave function networks can become scale free  - a mathematical structure that has found wide-spread application in several fields, ranging from power distribution and internet networks to epidemics~\cite{albert2002statistical,dorogovtsev2003evolution,posfai2016network}.
We demonstrate this property using experimental snapshots obtained on a Rydberg quantum simulator operating with more than 100 atoms~\cite{Browaeys2020,scholl2020programmable} and with large-scale numerical simulations using neural quantum states~\cite{Carleo2017,Markus2019}. We then argue about its generic applicability to state preparation protocols, and discuss how other types of networks - Erdos-Renyi~\cite{bollobas2012graph} - can instead emerge if the resulting dynamics describes uncorrelated states. In terms of observables, required resources and applicability regimes, our approach is complementary to other methods aimed at fully characterizing quantum states via snapshots such as those based on classical shadows~\cite{Huang2020}, randomized measurements~\cite{elben2020mixed,elben2022randomized,Brydges2019probing}, and chaotic dynamics~\cite{arute2019quantum,Choi2023}. Its main distinctive features, that we elaborate upon below, are direct interpretability and straightforward scalability for strongly correlated, low-temperature states.

The correspondence between quantum simulator outputs and conventional network theory immediately enables a transfer of methods and concepts from previously disconnected fields. 
We leverage this connection to address two challenges in the field of quantum simulation.
Firstly, we show that we are able to characterize the complexity of the quantum simulator output by determining its Kolmogorov complexity - the accepted absolute measure of information content of finite objects~\cite{kolmogorov1963tables,li2008introduction}, that quantifies the (in-)compressibility of the quantum wave function information as contained in the snapshots. This allows us to demonstrate the emergence of critical behavior at the level of information complexity, directly probing at the wave function level the emergent simplicity dictated by renormalization group theory.

Secondly, we introduce a method to perform cross-platform verification of quantum simulators~\cite{eisert2020quantum,carrasco2021theoretical}.
The method is based on the full network information without the need to perform an exponentially increasing number of measurements for increasing system size, which is the case for generic cross-verification based on the density matrix~\cite{eisert2020quantum,carrasco2021theoretical}.
By means of the Epps-Singleton test~\cite{epps1986omnibus} we identify, with statistical significance, a time scale beyond which cross-verification falters due to experimental imperfections not covered by our theoretical description. In addition, we provide statistically rigorous bounds for previously observed time-delay effects, that demonstrate the capability of our methods to identify systematic effects that are invisible to low-order correlation functions. 
Beyond these two demonstrative tools, the quantum wave function networks introduced in this work provide a new generically applicable framework to probe and characterize the quantum many-body wave function accessible in a variety of atomic and solid state quantum hardware, solely requiring  in situ imaging of the many-body wave function.

\section{Wave function networks: theoretical framework}

In this section, we describe how data sets generated by a collection of wave function snapshots can be represented by a network structure with nodes and links.
For the sake of simplicity, we consider a many-body system composed of spin-1/2 degrees of freedom defined on a two-dimensional lattice: the approach can be straightforwardly generalized to continuum theories, as well as to different types of local Hilbert spaces.

\paragraph*{Snapshot data set. -} Each wave function snapshot, labeled by an index $j$, takes the form:
\begin{equation}
    X_j [w] = (s_{1}^j, s_{2}^j, ...s_{N}^j)
\end{equation}
where $s_{m}^j$ is the measured value of the spin at position $m$. $N$ is the total number of sites in the system, while $w$ are the external parameters related to the snapshot - in our case the Hamiltonian couplings. Each of these configurations corresponds to a single data point embedded in a data space 
whose embedding dimension is $N$. This is depicted in Fig.~\ref{fig:Scheme}(a) with the three examples of green, orange and blue dots.

The data set we are interested in is formed by the collection of all available snapshots:
\begin{equation}
    \vec{X}[w] = \{ X_j\} = \{ X_1, X_2, ..., X_{N_r}\}
\end{equation}
where $N_r$ is the number of available snapshots, that is, the number of realizations. The data set might, in principle, include repetitions - e.g., $X_l=X_f$ for some $l\neq f$ - in particular, at very small volumes. It is possible to take care of them, as we detail in  ~\footnote{This can be done  at the data structure level in several manners - either utilizing estimators based on discrete distances, or adapting estimators with extra dimensions.}. However, to simplify the remainder of the discussion here, we assume no repetitions are present. 

\paragraph*{From data sets to wave function networks. -} 
\label{secIIb}
We now discuss how to translate the wave function snapshot data sets into a network structure.
There are two key choices that have to be made: 
(i) the selection of a proper metric in the embedding space, that allows one to compute distances between data points; 
and (ii) a criterion to activate links between data points, based solely on their distances.

The choice of a proper metric is an important aspect of the approach. 
Taking inspiration from recent results in the context of classical and quantum statistical mechanics models, we use the Hamming distance \cite{Mendes2020,Mendes2021}. 
Given two configurations $X_i, X_j$, such distance counts the number of spins that are aligned differently and reads
\begin{equation}\label{eq:distance}
    d(X_i, X_j) = \sum_{p = 1}^{N} | s^i_p-s^j_p|.
\end{equation}
The statistics of Hamming distances are related to arbitrary rank correlation functions between local degrees of freedom (i.e., $s_k$)\cite{Mendes2020}. Hence, they are sensitive to short-range and long-range correlations alike, which justifies their use as a similarity measure to define links between nodes. Specifically, we define a (geometric) network from our data sets by adopting the following procedure:
\begin{enumerate}
    \item Each point $X_i$ in the data set represents a node. 
    \item If two nodes are at distance $d<R$, we draw a link.
    \item The distance $R$ is chosen in a way that is dependent on the number of samples taken and reflects the typical value of distances for a given set of external parameters $w$.
    In particular, we define $R$ as 
    \begin{equation}
     R = \left< r_c \right> = \frac{1}{N_r} \sum_{i=1}^{N_r} r_c(i),
     \label{cutoff}
    \end{equation}
    where $r_c(i)$ is the distance between point $X_i$ and its $c$-nearest neighbor.
\end{enumerate}
An essential aspect of our approach is the choice of a cutoff $R$ that avoids overcounting isolated nodes while keeping a non-trivial network structure (i.e., in which all the nodes are not simply fully connected).
In particular, we show that our main conclusions are independent of the choice of $R$ for a certain range of values of $c$ in Eq. \eqref{cutoff}.
We discuss this issue in detail in Appendix \ref{AppI}.

\subsection{Network representation and correlations}

At a naive level, one could expect that such WFNs simply reflect the intrinsic randomness of the wave function sampling - that is, they are ultimately generated by a Poissonian process. It turns out that this intuition is fundamentally incorrect. 

In order to underpin the relation between network representation and correlations, we start by schematically illustrating the above procedure in Fig.~\ref{fig:Scheme}(a) (i) - (iv). A graphical example of a network with spin-1/2 systems and cutoff radius equal to 1 (that is, only configurations differing by a single spin flip are connected) is depicted in Fig.~\ref{fig:Scheme}(b): there, the black circle represents the Neel state that is connected to several other states by a single spin-flip - and, thus, minimal distance - while it is not connected to any other states. This example allows us to intuitively connect physical properties to network ones: a wave function network that carries correlations will feature 'hubs', that is, few states with many connections, and a lot of states with few connections. Conversely, a random "infinite-temperature" state will likely feature a majority of states with an intermediate number of neighbors, and won't feature either hubs or states with very few links.

The simple picture defined above is, {\it per se}, not particularly informative: however, it crucially sheds light on the classes of networks we can expect depending on how correlated the system is. This description of correlated states is reminiscent of what happens in several classes of scale-free networks: 
 typically described by a scaling relation of the probability distribution $P_k$ associated to the number of connections $k$ of each node (or, as is more commonly called, the degree distribution), that follows a power law distribution
\begin{equation}
    P_k \propto k^{-\alpha}.
\end{equation}
Such a function monotonically decreases with $k$, and allows us to distinguish between the majority of nodes that have few links, and the minority of those that have many links (see Fig.~\ref{fig:Scheme}(b)). While the prominence of hubs seems to be mostly relevant to ordered states, it is in fact a property that is even more robust in the presence of very strong correlations - such as, e.g., those emerging at quantum critical points. Conversely, networks representing random states will not be scale free, and can be construed as Erdos-Renyi (ER) networks - where the probability $P_k$ of a node having $k$ neighbors is approximately given by a Poisson distribution~\cite{bollobas2012graph}.

We emphasize that in the many-body regime, the number of snapshots, $N_r$, available from an experiment, is typically insufficient to tomographically reconstruct the wave function, i.e., $N_r \ll 2^N$.  
The WFN construction aims at a characterization of a state that focuses solely on the most important (yet unknown) degrees of freedom in the system, and not its entire data structure. So, our method is conceptually different from tomographic methods, including those based on specific ansatze.

\subsection{An illustrative example: quantum Ising model at equilibrium}
\label{sec:QIM}
\begin{figure}[t]
	\includegraphics[width=0.97\columnwidth]{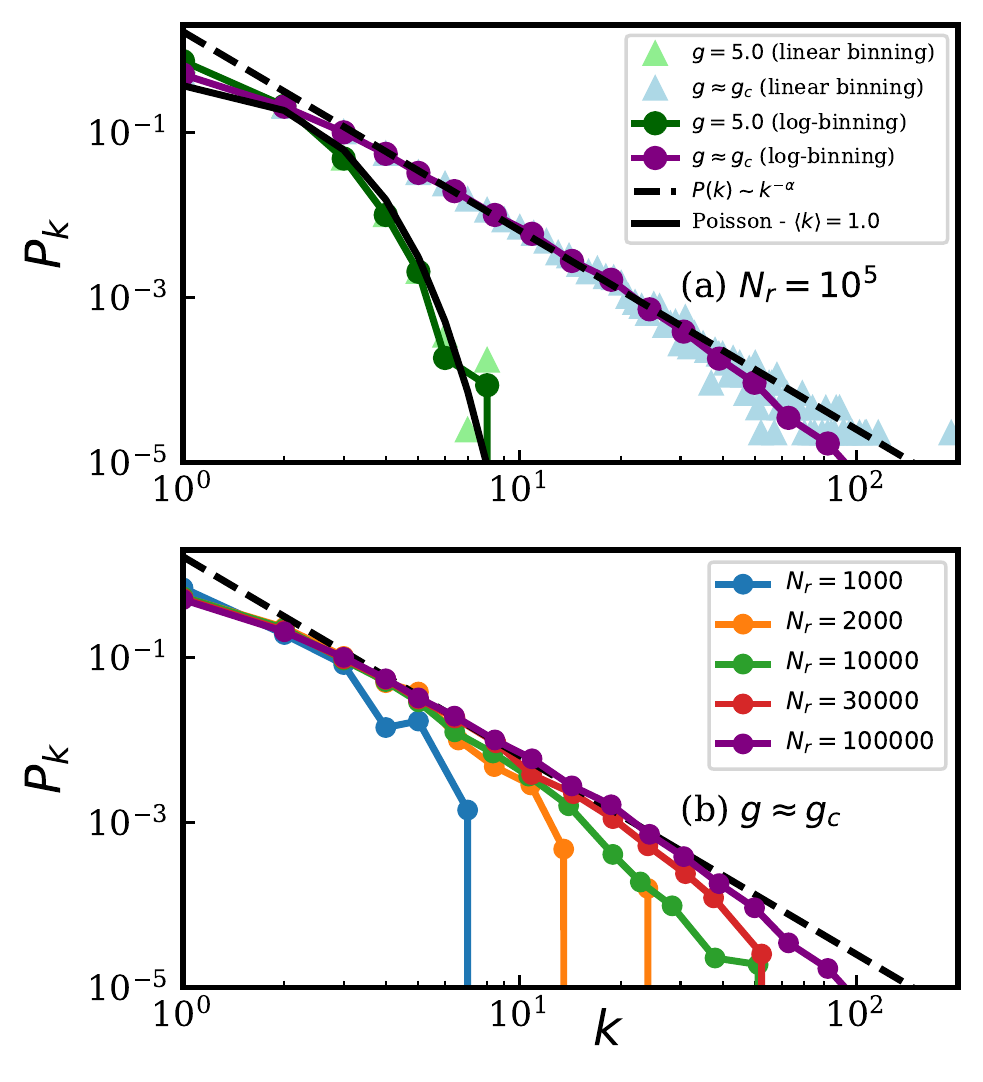}
	\caption{{ Degree distribution, $P_k$, for the WFN of the ground-state quantum Ising model.  Panel (a) shows  $P_k$ of the WFN with $N_r=10^5$ nodes for $g= 5.0$ and $g = 3.04 \approx g_c$. In the paramagnetic region, the resulting network is compatible with a Poisson distribution (solid line, i.e., Erdos-Renyi network)  with $\langle k\rangle =1$. As expected, in the vicinity of the critical point, the WFN becomes scale free, with $\alpha \simeq 2.4$ (dashed line). For comparison we compute $P_k$ using both linear (triangles) and logarithmic (circles) histograms.
 Panel (b) shows $P_k$ for different values of $N_r$ for $g \approx g_c$ using logarithmic histograms, again with the scale free ditribution shown as a dashed line.  In all cases we build the network using a cutoff $R = \left< r_1 \right>$ (where $\left< r_1 \right>$ is defined in Eq. \eqref{cutoff}).
	}
}
	\label{fig:Qising}
\end{figure}

Before discussing the experimental relevance of WFNs, we illustrate the emergence of scale-free networks in many-body systems by utilizing an example borrowed from equilibrium statistical mechanics. In Fig.~\ref{fig:Qising}, we show the degree distribution $P_k$ obtained via sampling the partition-function snapshots of the 2D quantum Ising model on a square lattice, with samples in the $z$-basis. The Hamiltonian reads:
\begin{equation}
    H = - \sum_{{i,j}} \sigma^z_i\sigma^z_j - g\sum_j \sigma^x_j.
\end{equation}
It features a quantum phase transition at $g_c \approx 3.04$ separating a non-correlated disordered phase (for $g > g_c$) from a ferromagnetically ordered state.
The corresponding $P_k$ is obtained by taking snapshots of the partition function, calculated via stochastic series expansion Monte Carlo simulations \cite{Sandvik1991,Sandvik2003} for a system of $N=L\times L= 8 \times 8$ sites, at inverse temperature $\beta = 2L$, which in our calculations was high enough to observe convergence within statistical uncertainty of energy and squared magnetization, i.e., to reach the ground state regime. Hence, the generated data sets correspond to the ground-state WF snapshots described above.

Fig.~\ref{fig:Qising} (a) displays the results from $N_r=10^5$ realizations. Deep in the paramagnetic phase, $g=5.0$, there are only weak correlations: the corresponding network is very well described by a Poisson distribution with $\langle k \rangle =1$. In the correlated regime, which is also the most entangled one, close to the phase transition, the network is described by a scale-free structure. We note that such a scale-free structure is unrelated to the absence of scale at criticality (scale-free networks can still be compatible with the presence of real-space finite scales~\cite{albert2002statistical}).

Once the degree $k$ of neighbors becomes a sizeable fraction of the total size of the network, we observe deviations from a scale-free profile, as expected. Above this size, the network properties are influenced by limited sampling~\cite{Caldarelli2021}. To further inspect this, we plot in the lower panel $P_k$ against $k$ for various $N_r$. We observe that indeed, the origin of the bending is due to the finite number of  samples, and that the curves for various $N_r$ are all compatible with a single power law, in this case, with exponent $\alpha\simeq 2.4$.
Changing the cutoff distance $R$ used to build the WFN does not affect the power-law scaling behavior of $P_k$ for $k$ above a certain threshold $k_c$; see Appendix \ref{AppI} and \ref{AppII} for more details.

At equilibrium, we expect the same dichotomic structure to appear generically in models that feature both weak- and strong-correlated regions. 
The key question we address below is, whether such structures are purely theoretical constructions, or if they can indeed be representative of the intricate dynamics taking place in quantum simulators, that are (i) off-equilibrium, open, and - a key difference from simulations - (ii) inherently probed with very high but not 100\% fidelity.

\begin{figure*}[t!]
	\includegraphics[width=1.0\textwidth]{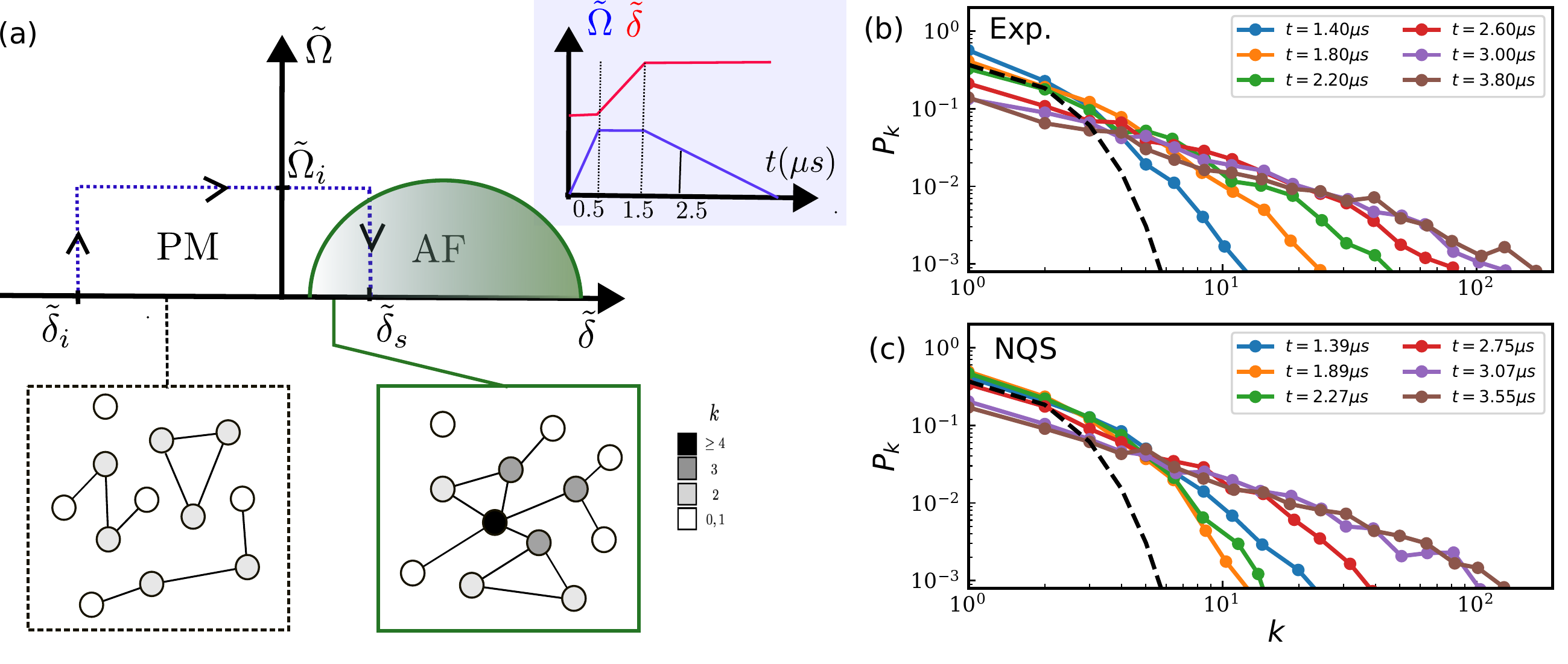}
    \caption{\textit{Observation of scale-free wave-function networks in Rydberg quantum simulators.} Panel (a) shows  a schematic  ground state phase diagram and the quasi-adiabatic state preparation scheme: the inset shows the sweep shape and the corresponding trajectory  is represented by the dashed lines in the phase diagram. In the paramagnetic (PM) regime, one expects a network description compatible with a Renyi-Erdos network, while in the vicinity to the antiferromagnetic (AF) region, that contains the Kibble-Zurek regime, a scale-free network structure is expected with power-law degree distribution, $P_k$, as illustrated by the network structures.
    The panel (b) presents the $P_k$ vs $k$  of the experimentally observed wave-function networks for a square lattice with $L = 8$. At short times, i.e., before crossing the phase transition, the distribution decreases exponentially (similar to Erdos-Renyi degree distribution with $\left<k\right>\simeq 1$, represented by the dashed lines in the graphs). At later times ($t>3~\mu s$), we observe a power law decay over two orders of magnitude, limited only by a bending that is due to a finite value of $N_r$. 
	Graph (c) shows NQS simulations of this quasi-adiabatic protocol for the same square lattice. The scale-free behavior of $P_k$ is again observed until one is sensitive to the effects of finite sampling. We note that the value of the decay exponent is $\alpha < 2$, signifying very stable wave function network properties, that will be discussed later in the presence of defects. For all cases we considered WFNs with $N_r = 2500$ nodes. 
	}
	\label{fig:ExpNet}
\end{figure*}

\section{Experimental observation of Erdos-Renyi and scale-free wave function networks}

\subsection{Experimental data and analysis of the network}

We now discuss the network structure of quantum simulation experiments. We analyze a recent experiment, that focuses on the quasi-adiabatic state preparation of a large antiferromagnetic state using a Rydberg quantum simulator \cite{scholl2020programmable}. This protocol plays a fundamental importance in quantum simulation and computing, and is very widely employed in atomic physics platforms. %\ab{Actually quenches are also popular...;-) } 
In addition, it typically features both regimes of no-correlations (short times), and of strong correlations, enabling us to test predictions based on both Erdos-Renyi and scale-free networks. Below, we summarize the main features of the experiment, that have been reported in \cite{scholl2020programmable}. 

The experiment consists of arrays of laser-cooled Rb atoms, individually trapped into optical tweezers separated by a distance $a$. Each atom can be considered as a pseudo-spin, the ground state being $\ket \downarrow$ and a  Rydberg state state being $\ket \uparrow$. Initially all the atoms are  prepared in $\ket \downarrow$. The atoms are then laser-excited to Rydberg states via a two-photon transition, so that the effective time-dependent Hamiltonian describing the dynamics reads:
\begin{equation}
    H(t) = \hbar\delta(t) \sum_i n_i + \frac{\Omega(t)}{2}\sum_i  \sigma^x_i+ \sum_{ij}J_{ij} n_in_j
    \label{eq:rydberg_hamiltonian}
\end{equation}
with $n_i = (\sigma^z_i +1)/2$, and $\sigma^\alpha_i$ Pauli matrices at the site $i$. Here, we have that $J_{ij}=C_6/r_{ij}^6$ as the atoms interact via the van der Waals interaction.
This quantum spin model exhibits both paramagnetic and antiferromagnetic phases in its ground state, for a schematic phase diagram see Fig.~\ref{fig:ExpNet}(a).
In the experiment a dynamical process has been implemented which, upon varying slowly $\Omega(t)$ and $\delta(t)$ over time, transforms an initial paramagnetic state into an antiferromagnetic one, as depicted in Fig.~\ref{fig:ExpNet}(a).
The adiabatic theorem guarantees that such a transformation is possible for ground states of systems with a nonzero gap whenever the parameter variations are sufficiently slow.
Close to a continuous quantum phase transition, however, the gap closes for a thermodynamically large system and excitations are generated unavoidably.
Importantly, the celebrated quantum Kibble-Zurek mechanism (QKZ) predicts that this defect generation, and on a more general level the dynamical properties itself of crossing such a transition, displays universal behavior controlled by the underlying quantum phase transition~\cite{Polkovnikov2005, zurek2005dynamics, Dziarmaga2005}.
In the context of two-dimensional systems, this has been recently described at the theoretical level~\cite{Markus2022}, and signatures have been observed in Rydberg experiments~\cite{ebadi2020quantum}.
For a finite-size system, such as the ones we deal here, the gap remains always finite. Because of this, a crossover from a QKZ regime towards an adiabatic regime emerges upon lowering the velocity of the ramp~\cite{Markus2022}.
In the experiment an antiferromagnetic ordering pattern has been achieved with a correlation length of the order of the system diameter, so that it is to be expected that the system resides in the crossover regime between QKZ and adiabaticity.

In what follows we will support the experimental data with numerically exact theory calculations, which will be key in a later stage in the cross-certification of the quantum simulator output.
For that purpose we will use neural quantum states (NQSs), which have been recently introduced as novel class of variational wave functions for the quantum many-body problem~\cite{Carleo2017}.
Most importantly for the purpose of this work, recent paramount advances have pointed out a route to numerically calculate quantum many-body dynamics in interacting two-dimensional quantum matter beyond what is achievable with other state-of-the-art methods~\cite{Markus2019,Markus2022}.
For details on the utilized numerical method we refer to Refs.~\cite{Markus2019,Markus2022} and to Appendix C.

Contrarily to the work in \cite{scholl2020programmable} we consider for our network analysis here two types of data sets: in the first one we use post-selected data without any defects in the array, i.e., each  trap  contains exactly one atom. 
In the second one we instead consider data sets including a mean number of defects of $\sim 3\%$, coming from an imperfect assembly of the atomic array \cite{Barredo2016}. The purpose of this second choice is that it will allow us to make quantitative statements on the resilience of scale-free structures, and most importantly, on their significance in terms of information - and, thus, complexity - content. 

{\it Scale-free and Erdos-Renyi networks. - } In Fig.~\ref{fig:ExpNet}(b), we plot the distribution $P_k$ for defect-free experimental data for square lattices of size $8\times 8$ and $N_r=2500$ at different times.
We identify two regimes: 

(A) At short times $t=1.52 \mu$s, $P_k$ decays exponentially with $k$, and its distribution resembles the one of a random ER network with $\langle k \rangle\simeq1$. This indicates that only limited correlations in the $z$-basis measurements are present in the system.

(B) Upon approaching the quantum phase transition ($t \sim 2.6 \mu$s) and at later times, the distribution changes drastically. In particular, we observe the emergence of a stable power-law profile with $\alpha < 2$ over almost two orders of magnitude, until at large $k$ finite sampling with $N_r < \infty$ introduces an inevitable cutoff in the form of an exponential decay. 
This phenomenology is characteristic of scale-free networks. 

In Fig.~\ref{fig:ExpNet}(c) we include as a comparison numerically exact theoretical results for $P_k$ by means of NQS simulations.
We utilize the same system parameters and number of samples as for the experimental data.
The simulations capture the exact same qualitative pattern described by the experiment, already indicating that, for the depicted timescale of the experiments, the effect of dissipation on the full many-body wave functions are likely to be negligible, and validating the microscopic modelling at a quantitative level.

\begin{figure}[t]
   \includegraphics[width=0.9\columnwidth]{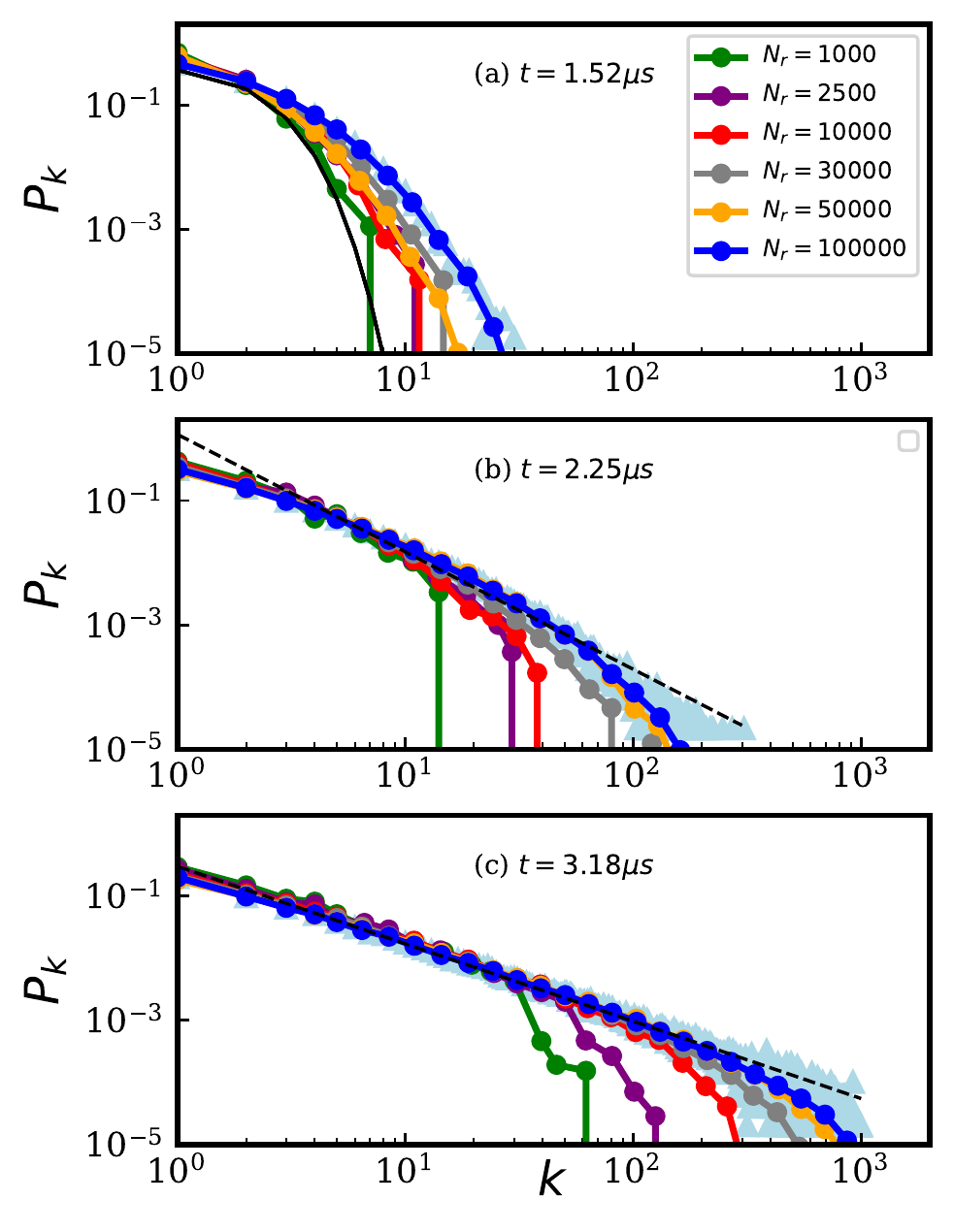}
    \caption{Dependence of the degree distribution, $P_k$, with the total number of nodes in the WFNs, $N_r$,  at the different values of $t$. In the scale-free regime the maximum size of the WFN at $k_{max}$ exhibit a strong dependence with $N_r$.
    The WFNs are obtained with data sets generated by NQS simulations of Rydberg experiments. }
    
	\label{fig:NQSNr}
\end{figure}

As depicted in Fig.~\ref{fig:ExpNet}, at large $k$, deviations from a power-law scaling become appreciable.
Such eventual deviations appear to be a sole effect of working with a finite number of samples $N_r<\infty$, which in turn implies that the range of the power-law behavior in $P_k$ can be extended by increasing $N_r$.
We show in Fig.~\ref{fig:NQSNr} the distribution $P_k$ for three reference cases of times $t$ by means of data obtained using NQS. Both qualitatively and quantitatively, $P_k$ exhibits the same features in all regimes: for ER graphs (Fig.~\ref{fig:NQSNr}(a)), increasing the number of nodes $N_r$ yields essentially the same structure of the network (keeping $\langle k \rangle\simeq 1$). For the case of scale-free networks, see Figs.~\ref{fig:NQSNr}(b,c), increasing the number of samples has the effect to enlarge the regime in $k$ of power-law behavior shifting the eventual bending, i.e., the deviation from the scale-free structure at large $k$, to larger and larger $k$.

\begin{figure}[b]
	\includegraphics[width=0.9\columnwidth]{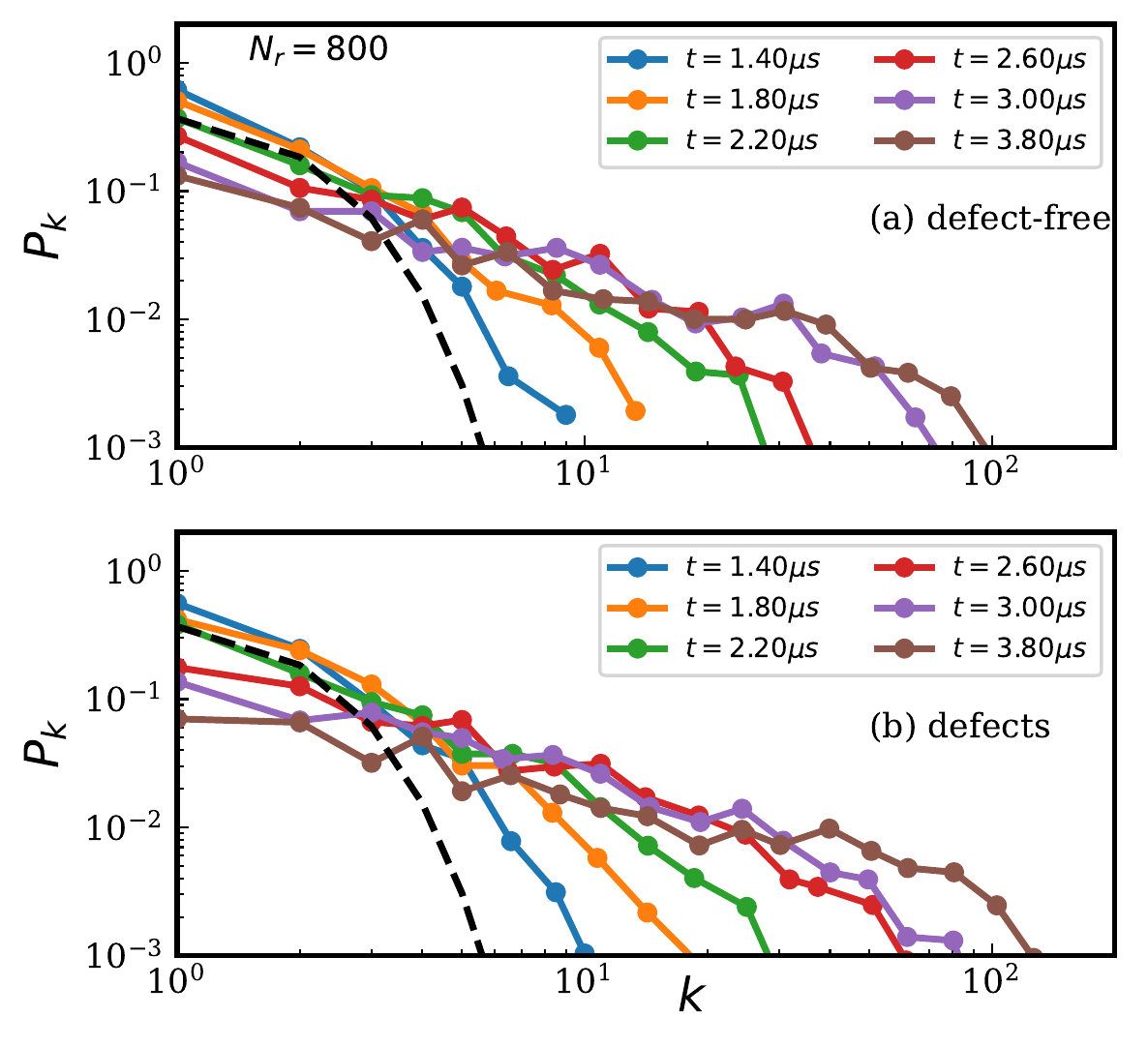}
	\caption{{\it Robustness of quantum simulator outputs.- } {Comparison of the degree distribution, $P_k$, of experimental WFNs generated without defects, and with a mean density of defects of $\sim 3\%$. We consider $N_r = 800$ in both cases. The results are qualitatively equivalent. 
	}
}
	\label{fig:Robust}
\end{figure}

{\it Robustness of quantum simulator outputs.- } We observe that at late times $t > 3 \mu s$, the exponent $\alpha$ of the power-law tail in $P_k$ satisfies the condition $\alpha < 2$ (see Fig.~\ref{fig:NQSNr} (c)).
As is known from network theory, scale-free networks with such an exponent $\alpha$ exhibit very robust information content with respect to perturbations.
We identify such a robustness also in the experimental data.
Specifically, as can be seen in Fig.~\ref{fig:Robust}, the experimental data sets with defects in the atomic array capture the same scaling behavior as without defects. In analogy to network theory, this analysis provides an interesting tool to characterize the robustness of quantum simulators based solely on their outputs, whenever they are described by scale-free or ER networks. An important comment is in order: such small values of the power law exponents are typically characteristic of finite networks: this is compatible with our theory, since we know that, in the infinite sampling limit $N_r\to\infty$ our network becomes infinitely large and it will be unavoidable to generate repetitions of the same snapshot. Such repetitions, however, we have excluded from the beginning and would require an adaption of our approach by means of certain weighted networks.

\subsection{Theory of wave function networks evolution over quasi-adiabatic state preparation}

The scale-free and ER WFN phenomenologies we observe in both experiment and numerical simulations are not tied to the specific problem we explore here, but, as we argue in this section, are generic features of quasi-adiabatic state preparation protocols.
Starting from an uncorrelated product state, it is natural to expect that at short times one may typically find random networks of wave function snapshots, i.e., networks with ER type structures. An example of such an instance is the case covered in this work, where we start from a product state with spins aligned in the $z$-direction. At short times the unitary dynamics will generate some weak but noticeable superposition of other configurations with a few flipped spins, which we expect to appear similar to the presence of local fluctuations such as those caused by dissipation or thermal fluctuations. These are inherently random, and should therefore yield an ER network, with Poisson-like degree distribution. For $N_r\ll 2^N$, such a process is expected to generate a very sparse network with $\langle k \rangle\simeq 1$, due to the fact that the average distance between configurations is roughly a constant. 

Upon approaching the quantum phase transition, we observe the emergence of a scale-free network structure.
The basic mechanism behind this can be understood upon the inspection of the introduced metric in Eq.~\eqref{eq:distance}, which is used to impose the fundamental underlying structure on our data sets. The network structure, which we probe through $P_k$, is generated by correlations in distances between different snapshots. Only in the case where such distances are correlated is it possible to find a power-law distribution $P_k$ of nodes having a connectivity $k$. As we discuss in the following, these correlations in distances between nodes in the network might be linked to the real-space correlation length in the system. Upon entering the quantum phase transition regime the system develops a large correlation length of the order of the system diameter due to the almost adiabatic dynamics generated through the experimental protocol.

From previous works on data analysis of snapshot measurements, it is expected that such large real-space correlation lengths yield Pareto distributions, i.e., power-law distributions, of distance measures in the data set~\cite{Mendes2020,Mendes2021}. In this light our observations of a scale-free network structure in $P_k$ appears natural in particular because $P_k$ quantifies correlations between distances of network nodes. 
We note that the scale-free property of a scale-free network solely concerns $P_k$ - indeed, other network properties may carry information that reflects the presence of a finite correlation length.

Once the quantum phase transition regime is reached, the system is effectively described by a large real-space correlation length.
Let us note that the dynamical behavior of the considered quantum spin model is expected to be potentially much richer as compared to the mostly studied case of dimension $D=1$.
Upon entering the broken-symmetry phase, the system will eventually thermalize implying an infinite correlation length.
In analogy to classical systems, the temporal process of generating a long-range ordered state is typically associated by coarsening and phase-ordering kinetics~\cite{Chandran2012}, which also comes along with universal power-law behavior.
In turn this means that upon crossing the quantum phase transition it is expected that the correlation length grows further in time also deep in the broken-symmetry phase.
When linking large real-space correlation lengths with scale-free network structures, this would imply that the scale-free network structure might survive also beyond the quantum phase transition region.
This underlines the universal character of the data structure dynamics observed in the experiments. The reasoning above applies also to first order phase transitions, as long as the correlation length at the critical point is larger than the system diameter (so that, in fact, the correlation functions in the system are not able to discern differences with respect to a continuous transition).

\section{Application 1: Kolmogorov complexity of wave function snapshots}

The output of a quantum simulator obtained via wave-function snapshots is, {\it per se}, a classical object. How complex must a classical computer program be, in order to reproduce such output? This is quantified by the so-called Kolmogorov Complexity (KC)~\cite{kolmogorov1963tables,li2008introduction}. 

For generic strings, computing the KC is a NP-hard problem. The same holds true for generic graphs, where the KC is quantified by the Haussdorf dimension~\cite{staiger1993kolmogorov}. This implies that computing the Kolmogorov Complexity of wave function snapshots is an extremely challenging task that cannot be undertaken in
general. 

However, as noted in the previous sections, quantum simulators often generate scale-free networks: for these, there exist known non-parametric learning algorithms that allow us to estimate the intrinsic dimension of the data points, and thus, the KC, in a manner that does not depend on scale. In particular, we utilize the 2-NN algorithm~\cite{Laio2017,glielmo2021unsupervised}, that has already been applied in the determination of critical properties of both classical and quantum statistical mechanics partition functions~\cite{Mendes2020,Mendes2021}.

\begin{figure}[t]
	\includegraphics[width=1.0\columnwidth]{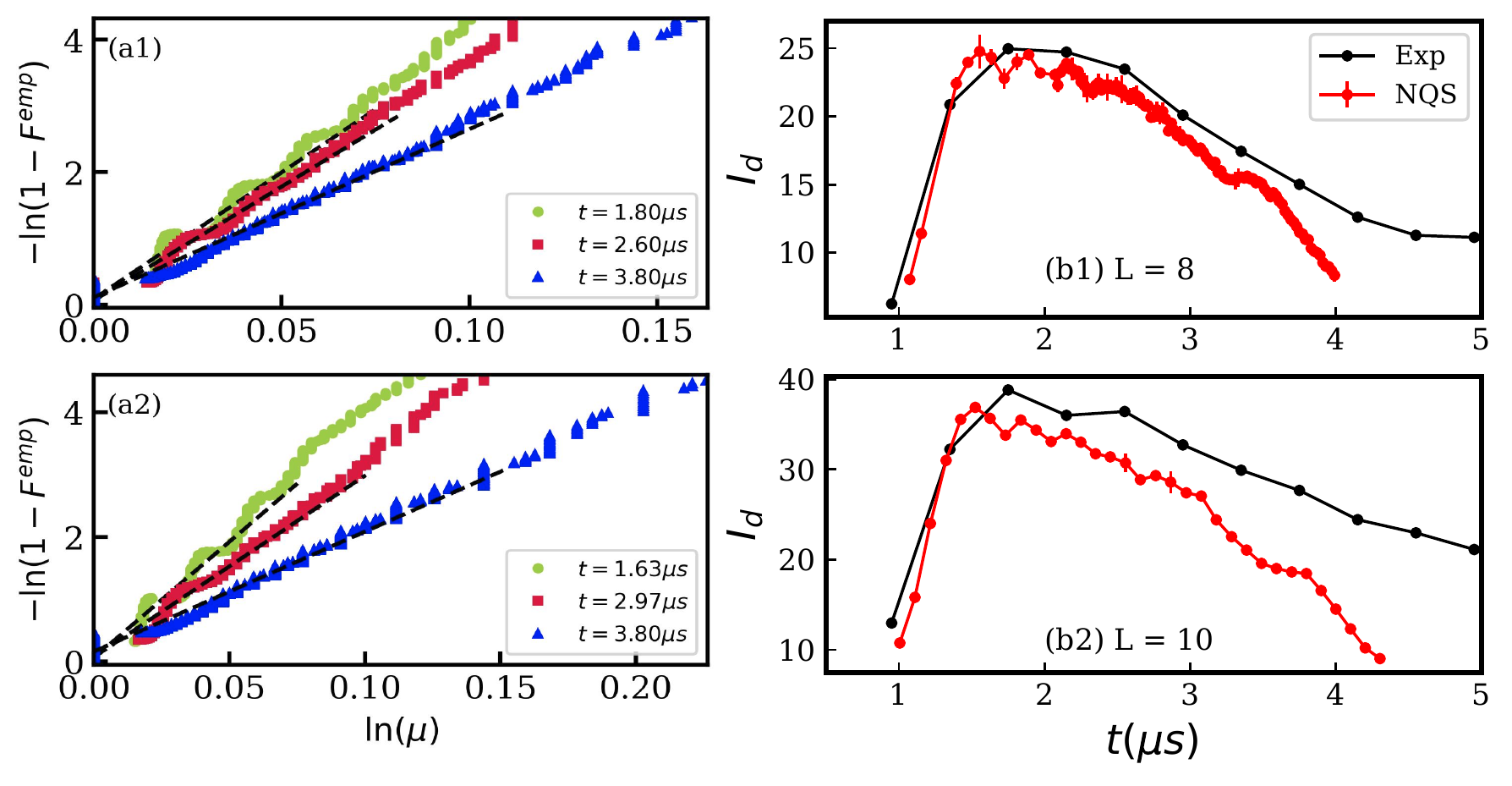}
	\caption{Complexity scaling in quantum simulators. Panels (a1-a2): cumulative distributions against $\mu=r_2/r_1$ for selected times.
	Panels (a1) and (a2) show results for experimental and NQS-simulations data sets, respectively.
	The quality of a description in terms of Pareto distribution (lines
	) increases as a function of time, for both simulation and experiment. Panels (b1-b2): time dependence of the intrinsic dimension, $I_d$, along the quasi-adiabatic time evolution for both experimental and simulated data sets. For $t > 2\mu s$, the complexity of the WFN is a monotonously decreasing function of time in both experiments and simulations
	, capturing the emergent simplicity (decrease of degrees of freedom) that is expected from the emergence critical behavior.  
	}
\label{fig:KC}
\end{figure}

The starting point is to consider, for each point $X_j$ in our data set, the distances to its first and second nearest-neighbor, $r_1(X_j)$ and $r_2(X_j)$, respectively. Under the condition that the data set is locally uniform in the range of second nearest-neighbors, it has been shown in Ref.~\cite{Laio2017} that the cumulative distribution function $F^{\text{emp}}$ of ${\mu = r_2(\vec{x})/r_1(\vec{x})}$ obeys:
\begin{align}
 I_d = - \frac{\ln\left[ 1 -  F^{\text{emp}}(\mu) \right]}{\ln\left( \mu \right)},
 \label{Id}
\end{align}
where $I_d$ is the intrinsic dimension of the data set. The intrinsic dimension quantifies the number of degrees of freedom required to capture the information content of the data set. While this is in principle a length-scale dependent property, our estimator directly focuses on the physically relevant distance that is determined by the sampling of the many-body wave functions.

In Fig.~\ref{fig:KC}(a), we depict the relation between $F^{\text{emp}}$ and $\mu$ obtained from (a1) experiments and (a2) NQS-simulations. In both cases, and for all times considered, the distribution is compatible with Pareto (additional oscillations appear at short times, likely due to the very simple structure of the network). These results guarantee the applicability of the 2-NN approach~\cite{Laio2017}.

\begin{figure*}[]
	\includegraphics[width=2.0\columnwidth]{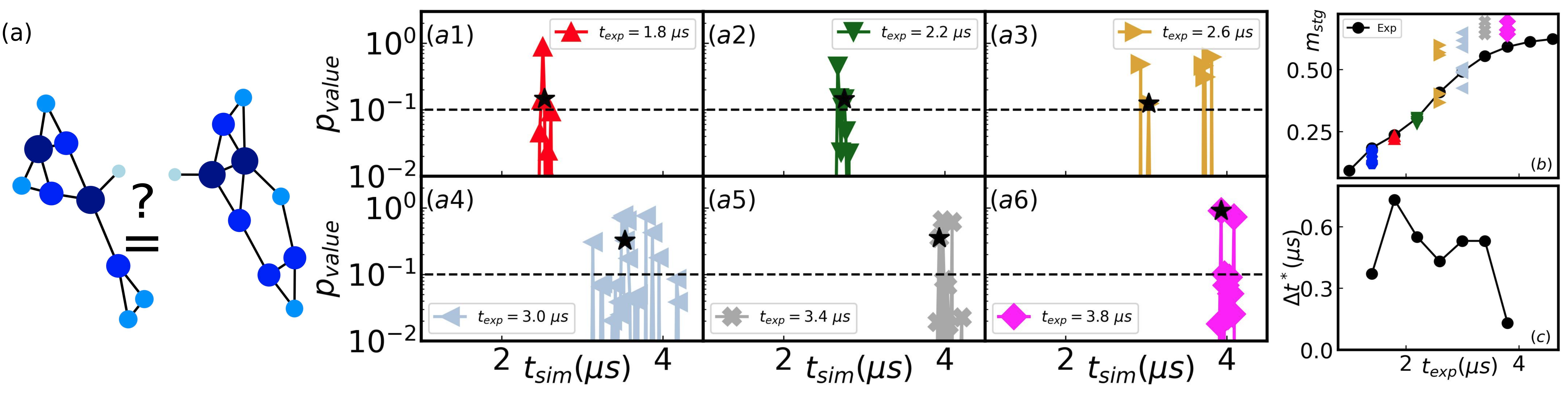}
	\caption{{Comparing the experimental and simulated WFNs, as illustrated in panel (a). In particular, we consider the Epps-Singleton two-sample test to check the hypothesis that the experimental and simulated degree distribution, $P_k$, are equal. For each experimental time $t_{\rm exp}$, we consider ES tests with the different simulated results at times $t_{\rm sim}$. Both WFNs have $N_r = 2500$ nodes, and we choose a cutoff distance $R = \left< r_1 \right>$ to generate them.
Panel (a1-a6) shows the corresponding $\pv$ as a function of $t_{\rm sim}$: results with $\pv > 0.1$ (marked by the dashed lines) are interpreted as statistically significant.
To cross-check our analysis, we also consider in panel (b) the order parameter, $m_{\rm stg}$, as a function of $t_{\rm exp}$. Each simulated result corresponds to the different times $t_{\rm sim}$ for which $\pv > 0.1$. Such an analysis allows us to identify $t_{\rm sim}^*$, the times where the best agreement exists between simulation and experimental data; the results corresponding to $t_{\rm sim}^*$ are marked as the black star points in the panels (a1-a6).
Finally, panel (c) shows the corresponding time-shift, $\Delta t^* = t_{\rm sim}^* - t_{exp}$, between experiments and NQS simulations (see text).}}
	\label{fig:ES_test}
\end{figure*}

In Fig.~\ref{fig:KC}(b), we show the time-dependence of the KC as measured by the intrinsic dimension across the ramp. Both experimental and simulation data clearly display two regimes: (i) up to $2~\mu$s, the complexity increases. This effect is trivial: the initial state is very close to a product state along the $z$-direction, so that at short times there is just one single dominant snapshot as the measurement outcome. The unitary evolution will necessarily generate additional correlations afterwards, thus increasing complexity. (ii) From $2~\mu$s onward, the complexity becomes a monotonously decreasing function of time. This second regime is a manifestation of the emergence of universal behavior whilst crossing a phase transition. Following quasi-adiabatic dynamics, the correlation length monotonously increases as a function of time: this implies that, in order to describe network properties, fewer variables are actually required - at equilibrium, these would just be the critical exponents and the amplitudes of correlation functions. The observations above are thus a direct manifestation of the emergent simplicity associated to universality at critical points \cite{Mendes2021}, and represent, to the best of our knowledge, the first experimental demonstration of the link between complexity and quantum critical behavior. 

We note that, after some time, NQS simulations predict a faster decrease of complexity with respect to the experimental data. We attribute this to the fact that the simulations can only partly keep track of the time evolution: the neural network structure utilizes a smaller number of effective variables, compatible with a decrease of KC, with respect to those that are describing the time evolution realized in experiment.

\section{Application 2: Cross-certification based on network properties}
\label{secV}
One of the key challenges for quantum computers and simulators is to verify their correct functioning or to certify the validity of their outcome. One basic idea in the field is cross-certification, which consists of directly comparing the output of one quantum machine with another - either quantum or classical. Recent protocols based on random unitary circuitry, aiming to compare the full ground-state wave functions, have been experimentally demonstrated to be superior to tomographic methods~\cite{Elben2020,elben2022randomized}. However, resources still scale exponentially with system size, making the present methods inapplicable to large devices. 

Here, we take a complementary angle and focus on a comparison based on wavefunction snapshots that take into account
the maximum amount of extractable information with currently available resources.
At the formal level, our goal is to compare two distributions in the limit where $N_r\ll 2^N$, i.e. which is relevant to experiments exploring many-body problems (oppositely, for $N\simeq 10$, it is possible to reach by brute force the regime $N_r\simeq 2^N$~\cite{Elben2020}). 
Clearly, a configuration-by-configuration comparison sampled by two distributions is meaningless unless the states are very close to a product state; for generic states, the probability of sampling the same set of configurations will scale exponentially to zero with system size.

The network representation we use allows us to bypass this limitation. 
Specifically, we wish to compare two WFNs obtained either by two experiments or by an experiment and a simulation, see Fig \ref{fig:ES_test} (a).
For the concrete case considered here, let us point out that the numerical simulation by itself is a formidable challenge, which we target again by means of the NQS approach~\cite{Carleo2017,Markus2019,Markus2022}, see also Appendix C for details.
Finding and quantifying similarities between two networks is a problem largely explored in different applications of network theory and is particularly useful for data sets that cannot be distinguished by direct inspection or low-order correlations~\cite{barabasi2016}.
In our case, such comparisons between networks are directly related to the choice of metric used to define the WFN. 
For scale-free WFN, this is particularly suitable, as we are guaranteed to have chosen a metric distance capturing correlations in the system.

As a simple and efficient way to compare experimental and simulated WFNs, we check the hypothesis that the corresponding degree distributions are equal by employing a nonparametric test, known as Epps-Singleton (ES) test \cite{Singleton1986}.
The latter allows us to identify, with statistical significance, when two WFNs are different. In the following, since we will employ this test to establish the identity of two WFNs, we will take as statistically significant proof of our claim, cases in which the p-value of the test takes values $\pv > 0.1$, i.e., in which both experimental and simulation data are compatible with a common probability distribution.

The results are summarized in Fig.~\ref{fig:ES_test}.
In panels (a1-a6), we present the corresponding $\pv$ of the ES tests obtained by comparing experimental data at a given time $t_{\rm exp}$ with the simulation data over a given time window (i.e., $1<t_{\rm sim} < 4.2 ~\mu$s).
This allows us to identify time windows where the quantum and classical simulators can be cross-certified with statistical significance in terms of the maximum amount of information available from their wave function networks. 
Interestingly, we note that the cross-certification agreement occurs at time windows that are shifted from the actual experimental times presented in the legends of Figs.~\ref{fig:ES_test} (a1-a6).

The fact that the cross-certification agreement of quantum systems occurs at a time $t_{\rm exp}$ that is different from times considered in the simulations $t_{\rm sim}$ can be attributed to miscalibrations of the parameters of the Hamiltonian (e.g., $\Omega(t)$ and $\delta(t)$). Similar observations are found when comparing experiments with simulations of physical observables based on matrix product states \cite{scholl2020programmable}.
Although such miscalibrations do not affect the actual physics, quantifying the corresponding time shift is essential for the cross-validation of the quantum simulators.

In general, we find that the ES test can provide, for a given $t_{exp}$, multiple candidate simulation times $t_{sim}$ for which $\pv > 0.1$, see Fig.~\ref{fig:ES_test}.
In order to finally select across these multiple candidates, we perform a second test by computing for each of the candidates an independent second quantity.
Here, we consider the staggered magnetization $m_{stg} = \sum_{i_x,i_y} (-1)^{i_x+i_y} \sigma^z_{i_x; i_y}$, where we included the results for all the candidate simulation times (see Fig.~\ref{fig:ES_test}(b)).
Finally, we choose the candidate $t_{sim}^*$, in which the simulated results for the order parameter are closest to the experimental data.
As one can see from Fig.~\ref{fig:ES_test}, up to a time of $t_{exp} = 3.0~\mu$s, we find that this procedure is capable of cross-certifying the experimental and theoretical data within the achievable accuracy, which is limited, for instance, by the finite time grid of the theoretical data.
For intermediate times $3.0 \lesssim t_{\rm exp} \lesssim 4.0~\mu $s small deviations start to emerge, whereas for times $t_{exp} \gtrsim 4.0~\mu $s the cross-certification fails, which could be caused by dissipation effects in the experiment that are not included in the theory calculation, or by a decreasing accuracy of our variational computation (similarly to what is observed in the complexity scaling).

This scheme further defines an optimal time-shift $\Delta t^* = t_{sim}^* - t_{exp}$ for the experimental data.
Fig.~\ref{fig:ES_test} (c) shows the estimated values of $\Delta t^*$. Importantly, in the time interval that the quantum simulator can be cross-certified, we identify a small time dependence of $\Delta t^*$, which has not been addressed previously. We note that the procedure does not work well for $t<1.5\mu$s, as expected: there, the network is not scale-free yet, so a direct comparison can only provide some rough qualitative guidance.

\section{Conclusions and Outlook}

We have introduced a network theory framework to interpret the maximum amount of information extractable from quantum simulators - wave function snapshots. Remarkably, such networks can become scale-free for strongly correlated states of matter, and are of direct experimental relevance, as we demonstrate with data from a large-scale Rydberg atom array experiment. We have illustrated the power of network description with two applications: demonstrating the scaling of complexity across a quantum phase transition during Kibble-Zurek scaling, and cross-certifying the wave function of a quantum and classical simulator up to system sizes that have never been attained previously.

Our work opens up a series of research directions based on a transfer of methods and concepts between network and quantum science. At the big picture level, it would be important to determine to which extent Erdos-Renyi and scale-free networks are able to characterize quantum simulators and computers. While our framework provides strong evidence that it works at and close to equilibrium, the structure of wave function networks in genuinely out-of-equilibrium situations is presently completely unknown. Understanding network properties corresponding to such dynamics might provide qualitative insights into how equilibrium is established at the wave function level, complementing current efforts focusing on observables, and providing direct links between dynamics and Kolmogorov complexity. Going beyond the case of unitary dynamics, understanding the role of dissipation might help characterize the stability of quantum dynamics to noise, which will ultimately always kick in and - very likely - imprint an Erdos-Renyi structure onto the system wave function.

In addition to conceptual insights, our framework is ideally suited to developing scalable quantum information tools. Examples range from improving cross-certification methods, and, most relevantly, applying them to data sets from large scale experiments, for which computing direct wave function overlap is hopeless. On a broader level, we believe that the parallelism between two very active, but so far disconnected, fields could be an ideal playground for developing new insights into how information is associated to many-body phenomena.

\begin{acknowledgements}
We thank G. Bianconi, J. Grilli, M. Marsili, R. Panda, R. Verdel, V. Vitale and P. Zoller for insightful discussions.
The work of M.~D. and A. A. was partly supported by the ERC under grant number 758329 (AGEnTh), and by the MIUR Programme FARE (MEPH). M.~D. further acknowledges funding within the QuantERA II Programme that has received funding from the European Union's Horizon 2020 research and innovation programme under Grant Agreement No 101017733. D.B. acknowledges support from MCIN/AEI/10.13039/501100011033 (RYC2018-025348-I and NextGenerationEU PRTR-C17.I1).
This project has received funding from the European Research Council (ERC) under the European Union’s Horizon 2020 research and innovation program (Grant Agreement No. 853443).
Moreover, the authors gratefully acknowledge the Gauss Centre for Supercomputing e.V. (www.gauss-centre.eu) for funding this project by providing computing time through the John
von Neumann Institute for Computing (NIC) on the GCS
Supercomputer JUWELS at Jülich Supercomputing Centre
(JSC) \cite{jsc}. This work is supported by the European Union's Horizon 2020 research and innovation 
program under grant  agreement No. 817482 (PASQuanS), 
the Agence Nationale de la Recherche (ANR, project RYBOTIN), and the 
European Research Council (Advanced grant No. 101018511-ATARAXIA).
\end{acknowledgements}

\newpage

\section{Appendices}

\subsection{Wave Network structure and the choice of the distance cutoff $R$}
\label{AppI}
As described in Sec. \ref{secIIb},  the structure of the wave function network (WFN) is defined by choosing a cutoff distance, $R$, in an embedded space defined by the Hamming distances, which allow us to define links between nodes.
We now discuss in more detail the influence of $R$ on the observation that WFNs can exhibit a scale-free structure.

Let us consider the list of all pair of distances $d(X_i,X_j)$ between nodes $X_i$ and $X_j$.
One crucial aspect to consider is that
the choice of $R$ is  bounded by the minimum and maximum distances on such a list (let us call $d_{min}$ and $d_{max}$, respectively):
$R < d_{min}$ would generate a network with all the nodes isolated, while $R > d_{max}$ would generate a featureless, fully connected network.
The choice of $R$ introduced on Eq.~\eqref{cutoff} naturally takes into account the typical scale distance in the embedded space, which depends on the $N_r$ or the Hamiltonian parameters.

Another important aspect is that we deal with distances in a ``high-dimensional'' embedded space 
(the embedded dimension is equal to the number of spins, $N$), where the so-called curse of dimensionality is expected to play a fundamental role.
For instance, we could expect that the difference between the minimum and the maximum distance (i.e., $d_{min} - d_{max}$) would become indiscernible compared to any reasonable choice of $R$ \cite{Shaft1997} given that the volume of a high-dimensional space increases so fast that the available data becomes sparse when  $N_r \ll 2^{N}$.
If this were the case, we would have observed just a featureless, fully connected network.
In the correlated regime, however,  we observe non-trivial network structures, which can be attributed to the fact that, in reality, the intrinsic dimension of the WFNs is much lower than the dimension of the embedded space \cite{Mendes2020,Mendes2021}.

\begin{figure}[t]
	\includegraphics[width=1.0\columnwidth]{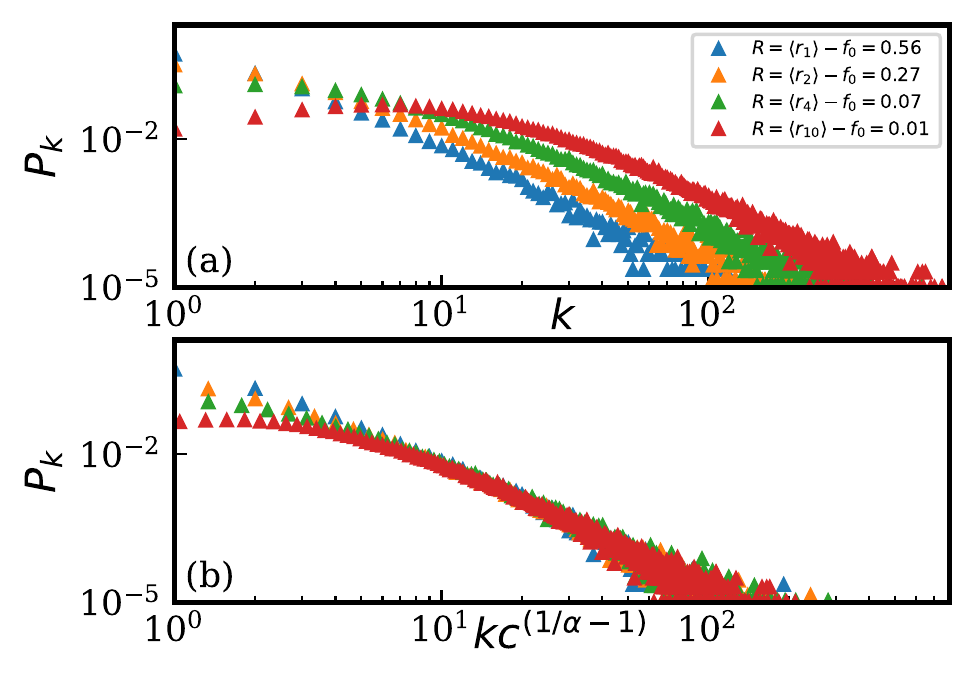}
	\caption{{ Degree distribution, $P_k$, for the WFNs of the ground-state quantum Ising model at the critical point, $g = g_c$.   Panel (a) shows  $P_k$ of the WFNs with different values of $R = \left< r_c \right>$ (see Eq. \eqref{cutoff}). We also show the fraction of isolated nodes, $f_{N_0}$. In panel (b) we consider a phenomenological collapse for the different $P_k$.
	}
}
	\label{figA1}
\end{figure}

Let us discuss how changes on $R$ influence the scale-free WFN.
Figure \ref{figA1} shows the degree distribution $P_k$ associated with the WFN generated at the quantum critical point of the quantum Ising model.
By increasing $R$, we observe two main effects. First, the $P_k$ is shifted by larger values of $k$. Second, the threshold $k_c$ above which $P_k$ starts to behave as power law increases, see Fig.~\ref{figA1} (a).
Specifically, we observe  that our data for different values of $R$ collapse in a same curve when we rescale the  $x$-axis; see Fig.~\ref{figA1} (b).
This result indicates that for the scale-free WFN, the main effect of increasing $R$ is to enlarge the cutoff $k_c$ below which the network is not scale-free.
In addition, we show the fraction of isolated nodes, $f_{N_0}$ by increasing $R$. 
For  $R = \left< r_{10} \right>$ less then $1\%$ of the nodes are isolated, however we still observe a power law behavior for almost one decade in $k$. 
Overall, we notice that for $g \approx g_c$ we can always observe a scale-free WFN for a wide range of choices of $R$.

\subsection{Power law fitting of the degree distribution}
\label{AppII}

One of our key results in this work is that wave function networks (WFNs) emerging in the vicinity of the quantum critical points are scale-free networks.
As we discuss in this section, our conclusion is based on the empirical observation that it is very likely that a power law function describes the corresponding degree distribution, i.e.,  $P_k \sim k^{-\alpha}$.

First, given a WFN, we count the number of links $k$ of each point in the network. The corresponding list of values is used to generate the histogram  $P_k$. Second, we fit the $P_k$ by employing the approach proposed in Ref. \cite{Clauset2007}. Specifically, in such an approach, one (i) defines an optimal scaling range, i.e., $k > k_{min}$, and the value for the power-law exponent $\alpha$  by selecting an optimal fit. In particular, the one that minimizes the Kolmogorov-Smirnov (KS) distance between the empirical data and a set of trial fits. (ii) The power-law distribution (i.e., with the selected $k_{min}$ and $\alpha$) is used to generate many synthetic data sets, which are compared with the power-law form by using the KS distance. The fraction of the synthetic KS distances that are larger than the empirical KS distance is then defined as the p-value of the test statistics. If the resulting p-value is greater than $0.1$, the power law is a plausible hypothesis for $P_k$; otherwise, it is rejected \cite{Clauset2007}. To implement the described approach, we use the Python power law package of Ref. \cite{Diermar2014}.

\begin{figure}[t]
	\includegraphics[width=1.0\columnwidth]{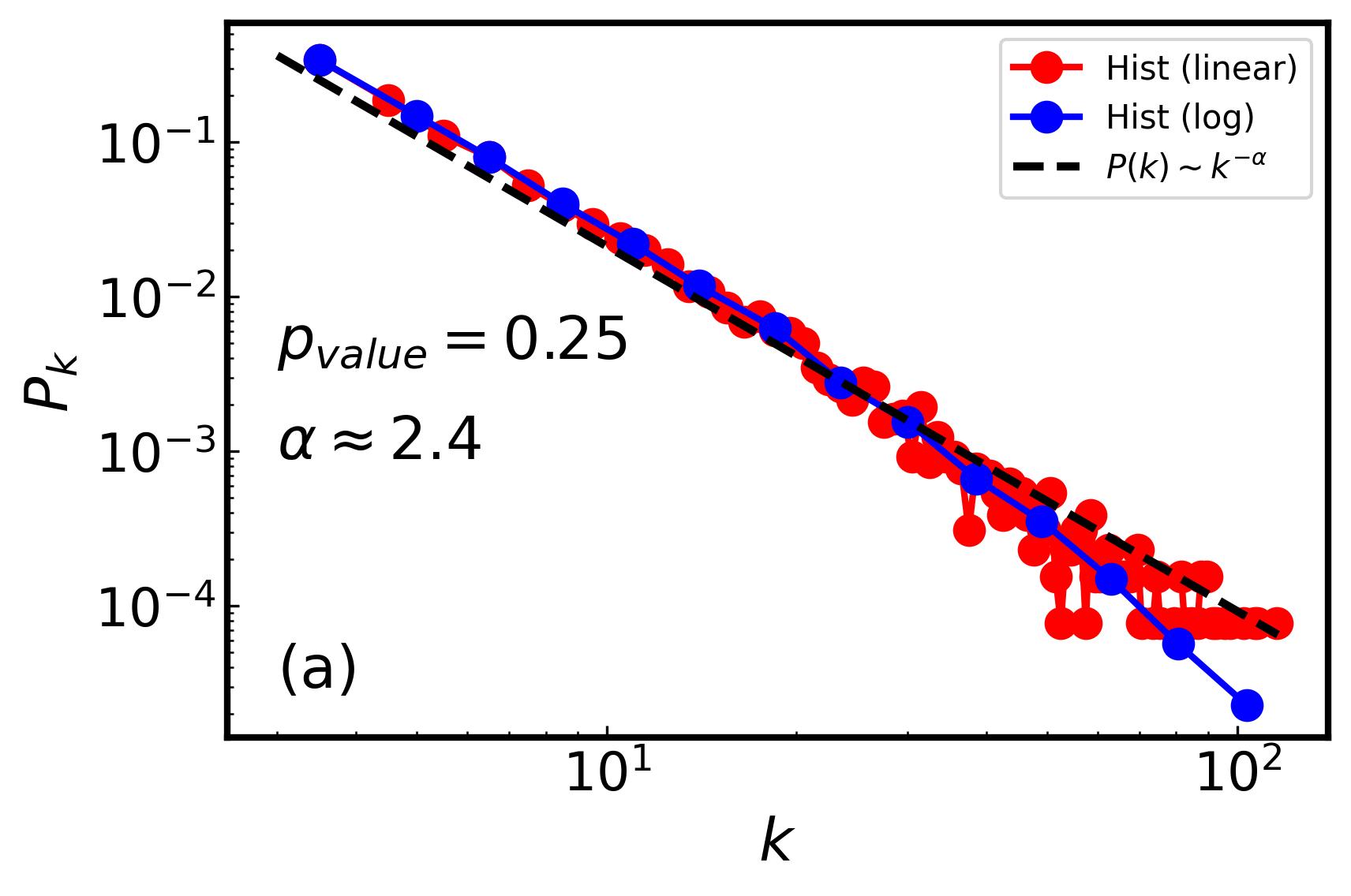}
	\caption{{Power law fitting of the degree distribution presented on Fig.~\ref{fig:Qising}. We use Komolgorov-Smirnoff statistics to perform the fitting and define the p-value (see text). 
	}
}
	\label{fig:PkD}
\end{figure}

As shown in Fig.~\ref{fig:PkD}, when considering a fitting of $P_k$ within an interval between $k_{min} = 3$ and $k_{max} = 120$, the results of the ES test satisfy the criterion $\pv > 0.1$.
We note that for obtaining $\pv > 0.1$ we have to impose a maximum cutoff $k_{max}$ for $P_k$.
The departure from the scale-free behavior for $k > k_{max}$ can be attributed to the finite size of the network, $N_r$.
Based on the analysis presented on Fig.~\ref{fig:Qising} (b), we indeed observe the range in $k$ that the power-law behavior is observed increases with $N_r$.
For future works, it will be interesting also to investigate the role of system size (i.e., $L = \sqrt{N}$) in the scale-free behavior of $P_k$. In particular, if it is possible to establish a finite-size scaling expression addressing the role of both the network-size $N_r$ and system-size $L$.

\subsection{Simulations with Neural Quantum States}
\label{AppIII}
Neural quantum states (NQS) have emerged recently as a new versatile class of variational wave functions \cite{Carleo2017}. The goal is to find an efficient representation of a many-body wave function $\ket{\psi}$ in the form of a parameterized function $\psi_{\vec\theta}(\vec s)$ that maps a computational basis configuration $\vec s=(s_1,\ldots,s_N)$ to a complex number, such that
\begin{align}
    \ket{\psi_{\vec\theta}}=\sum_{\vec s}\psi_{\vec\theta}(\vec s)\ket{\vec s}\ .
\end{align}
Here, $\ket{\vec s}=\ket{s_1}\otimes\ldots\otimes\ket{s_N}$ denotes the computational basis states of a system with $N$ degrees of freedom, and for our purposes $s_i\in\{\uparrow,\downarrow\}$.
There is a number of appealing reasons to choose $\psi_{\vec\theta}(\vec s)$ in the form of an artificial neural network (ANN) to render the ansatz a NQS. Most importantly, rigorous representation theorems guarantee that any possible wave function can be approximated by an ANN in the limit of large network sizes \cite{Cybenko1989,Hornik1991,Kim2003,LeRoux2008}. This means that the approach is numerically exact in the sense that the accuracy of results can be certified self-consistently by convergence checks. While the general function approximation theorems do not tell us whether the representation in the form of an ANN is efficient, it has been shown that NQS cover some volume law entangled states and correlated states of systems in two spatial dimensions, which are notoriously difficult to capture with established methods \cite{deng2017,Gao2017,sharir2019,sharir2022,roth2022highaccuracy,reh2023optimizing}. Finally, the complexity of the algorithms involved scales gently with system size and number of parameters, and large parts are amenable to large-scale parallelization to take advantage of distributed GPU clusters \cite{jvmc}.

While the variational ansatz with a limited number of parameters solves the problem of efficient representation, the efficient extraction of information from the wave function is achieved by Monte Carlo sampling. For example, the quantum expectation value of an operator $\hat O$ can be rewritten as
\begin{align}
    \bra{\psi_{\vec\theta}}\hat O\ket{\psi_{\vec\theta}}
    =\sum_{\vec s}\frac{|\psi_{\vec\theta}(\vec s)|^2}{\braket{\psi_{\vec\theta}}{\psi_{\vec\theta}}}O_{\text{loc}}(\vec s)
\end{align}
with the local estimator $O_{\text{loc}}(\vec s)=\sum_{\vec s'}O_{\vec s,\vec s'}\frac{\psi_{\vec\theta}(\vec s')}{\psi_{\vec\theta}(\vec s)}$ that can be computed efficiently for local operators with only a polynomial number of non-vanishing matrix elements $O_{\vec s,\vec s'}=\bra{\vec s}\hat O\ket{\vec s'}$. This means that the expectation value can be estimated efficiently by Monte Carlo sampling the Born probability distribution $\frac{|\psi_{\vec\theta}(\vec s)|^2}{\braket{\psi_{\vec\theta}}{\psi_{\vec\theta}}}$ \cite{van_den_Nest2009}, and the same holds for all quantities of interest appearing in NQS algorithms. Notice that the only way to access the wave function in quantum simulation experiments are projective measurements, which are likewise a sampling of the Born distribution; this is a very useful parallel when attempting a direct comparison of the obtained data, because obtaining samples from the wave function could turn out to be very costly with alternative numerical approaches \cite{scholl2020programmable}.

An optimal approximate solution of the Schrödinger equation $i\frac{d}{dt}\ket{\psi_{\vec\theta}}=\hat H\ket{\psi_{\vec\theta}}$ within the manifold of wave functions $\ket{\psi_{\vec\theta}}$ is obtained via a time-dependent variational principle (TDVP) \cite{BROECKHOVE1988547,Carleo2017,Markus2019}. This leads to an ordinary differential equation prescribing the time evolution of the variational parameters,
\begin{align}
    \text{Im}\big[S_{k,k'}\big]\dot{\theta}_{k'}=-\text{Im}\big[iF_{k}\big]
    \label{eq:tdvp}
\end{align}
with the quantum metric tensor 
$S_{k,k'}=
\braket{\partial_{\theta_k}\psi_{\vec\theta}}{\partial_{\theta_{k'}}\psi_{\vec\theta}}
-\braket{\partial_{\theta_k}\psi_{\vec\theta}}{\psi_{\vec\theta}}\braket{\psi_{\vec\theta}}{\partial_{\theta_{k'}}\psi_{\vec\theta}}$
and the force vector $F_k=\bra{\partial_{\theta_k}\psi_{\vec\theta}}\hat H\ket{\psi_{\vec\theta}}-\bra{\partial_{\theta_k}\psi_{\vec\theta}}\ket{\psi_{\vec\theta}}\bra{\psi_{\vec\theta}}\hat H\ket{\psi_{\vec\theta}}$; notice that the imaginary part appears on both sides of the equation as we are considering real parameters \cite{BROECKHOVE1988547,jvmc}.
Hence, the time-evolved wave function starting from a given initial state can be obtained by integrating Eq.~\eqref{eq:tdvp}. In previous works it was found that careful regularization is crucial to achieve state-of-the-art results in this way \cite{Markus2019,Markus2022}. For the present work we developed a new way of phrasing and solving the variational problem, which we call the \textit{conditional} TDVP. The details of this approach will be described in a separate manuscript \cite{ctdvp}. All results presented here were obtained in this way.

\begin{figure}
\includegraphics[width=0.9\columnwidth]{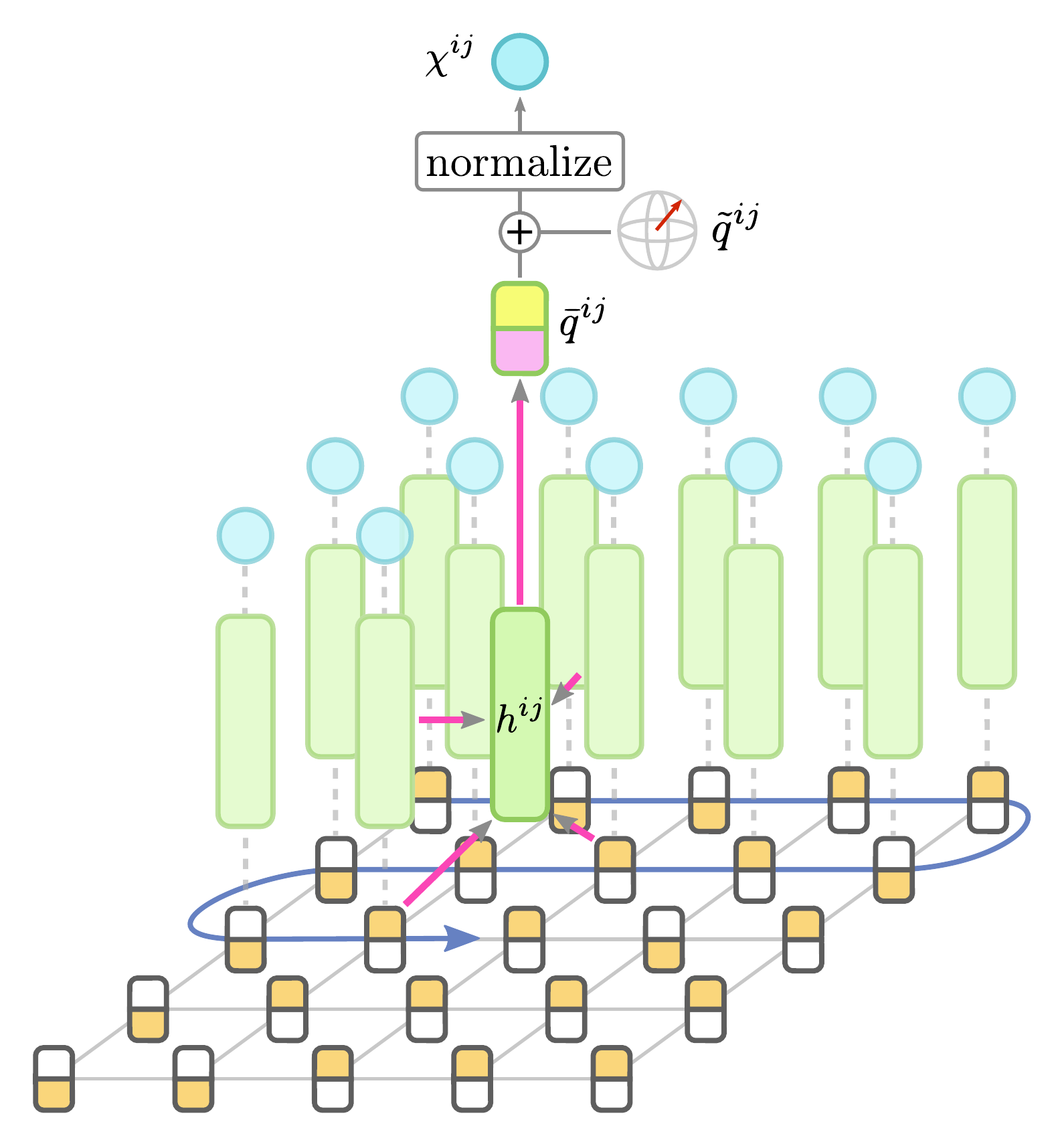}
\caption{Schematic depiction of the used neural network architecture. For evaluation the lattice is traversed along the path indicated by the blue arrow. A hidden state $h^{ij}$ is computed at each site using the one-hot encoded local basis configurations and the hidden states of previously visited neighboring sites as indicated by the pink arrows, which correspond to dense layers. From the hidden state a correlated contribution to the conditional qubit state, $\bar q^{ij}$, is computed and an additional uncorrelated contribution $\tilde q^{ij}$ is added to it to obtain the logarithmic conditional amplitudes $\chi^{ij}$ after normalization as noted in Eq.~\eqref{eq:nqs_final_step}.}
\label{fig:nqs_architecture}
\end{figure}

The network architecture used in our simulation is a variant of the recurrent neural network (RNN) for two-dimensional systems introduced in Ref.~\cite{hibatallah2020}. The structure of this architecture is depicted schematically in Fig.~\ref{fig:nqs_architecture}.
The starting point is a one-hot encoding $\vec\sigma_{i,j}$ of the local spin configurations $s_{i,j}$, i.e., $\vec\sigma_{i,j}=(1,0)$ if $s_{i,j}=\uparrow$ or $\vec\sigma_{i,j}=(0,1)$ if $s_{i,j}=\downarrow$.
The neural network is then evaluated by traversing the two-dimensional lattice in a snake-like manner. Let us denote the $k$-th lattice site index along the snake path as $(i_k,j_k)$ and assume that the linear dimension of the lattice is $L$.
At each lattice site, a conditional single qubit state $\psi(s_{i_k,j_k}|s_{1,1},\ldots,s_{i_{k-1},j_{k-1}})$ is generated in the way detailed below. From these conditional states, the coefficient of the many-body wave function is obtained as
\begin{align}
    \psi(\vec s)=\prod_{k=1}^{L^2}\psi(s_{i_k,j_k}|s_{1,1},\ldots,s_{i_{k-1},j_{k-1}})\ .
    \label{eq:ar}
\end{align}

 For the conditional states at every lattice site, a local hidden state $\vec h^{(i,j)}$ is computed based on the spin configuration and hidden state of two neighboring sites as
\begin{align}
    h^{(i_k,j_k)}_l&=f\big(W^{H}_{lm} h^{(i_{k-1},j_{k-1})}_m+W^V_{lm}h^{(i_{k-L},j_{k-L})}_m\big)\nonumber\\
    &\quad+f\big(W^{S_1}_{lm}\sigma^{(i_{k-1},j_{k-1})}_m+W^{S_2}_{lm}\sigma^{(i_{k-L},j_{k-L})}_m\big)\ .
\end{align}
Here, $f$ denotes the non-linear activation function and $W^{(\cdot)}_{lm}$ denote the weights of the dense layers; double indices indicate summation. At the boundaries, where required neighboring sites do not exist, the corresponding input is replaced by zeros. Next, the hidden state is processed by a dense layer with two-dimensional output $(\bar{q}^{ij}_{\text{R}}, \bar{q}^{ij}_{\text{I}})$, corresponding to the real and imaginary parts of a complex number $\bar q_{ij}$. This number constitutes the correlated contribution to the logarithmic $\uparrow$-coefficient of the conditional local qubit state, up to normalization and a global phase. In addition, we introduced one complex-valued variational parameter $\tilde{q}^{ij}=\tilde{q}^{ij}_{\text{R}}+i\tilde{q}^{ij}_{\text{I}}$ for each lattice site, which corresponds to a contribution to the conditional qubit state that is uncorrelated. With $q^{ij}=\bar{q}^{ij}+\tilde{q}^{ij}$ we finally produce the logarithmic conditional wave function amplitudes
\begin{align}
    \chi^{ij}_\uparrow &= \frac12\log\Bigg(\frac{\exp\big(q^{ij}_{\text{R}}\big)}{1+\exp\big(q^{ij}_{\text{R}}\big)}\Bigg)+iq^{ij}_{\text{I}}
    \nonumber\\
    \chi^{ij}_\downarrow &= \frac12\log\Bigg(\frac{1}{1+\exp\big(q^{ij}_{\text{R}}\big)}\Bigg)+iq^{ij}_{\text{I}}
    \label{eq:nqs_final_step}
\end{align}
such that
\begin{align}
    \psi(s_{i_k,j_k}|s_{1,1},\ldots,s_{i_{k-1},j_{k-1}})=\exp\big(\chi^{i_kj_k}_{s_{i_k,j_k}}\big)\ .
\end{align}

The uncorrelated contribution $\tilde{q}^{ij}$ extends the standard RNN architecture. We introduced it, because we found it difficult with the plain RNN to capture the initial part of the control protocol, where only the orientation of the uncorrelated qubit states is rotated and hardly any correlations are produced. In our architecture $\tilde{q}^{ij}$ can fully capture the product state, such that the job of the RNN is just to account for correlations on top of it. Including $\tilde q^{ij}$ does not affect the autoregressive property of the ansatz introduced by the decomposition into a product of conditionals \eqref{eq:ar}. This means that the architecture allows for direct sampling of uncorrelated configurations at the cost of a single network evaluation per sample \cite{sharir2020,hibatallah2020}.

\begin{figure}[b]
\center
\includegraphics[width=\columnwidth]{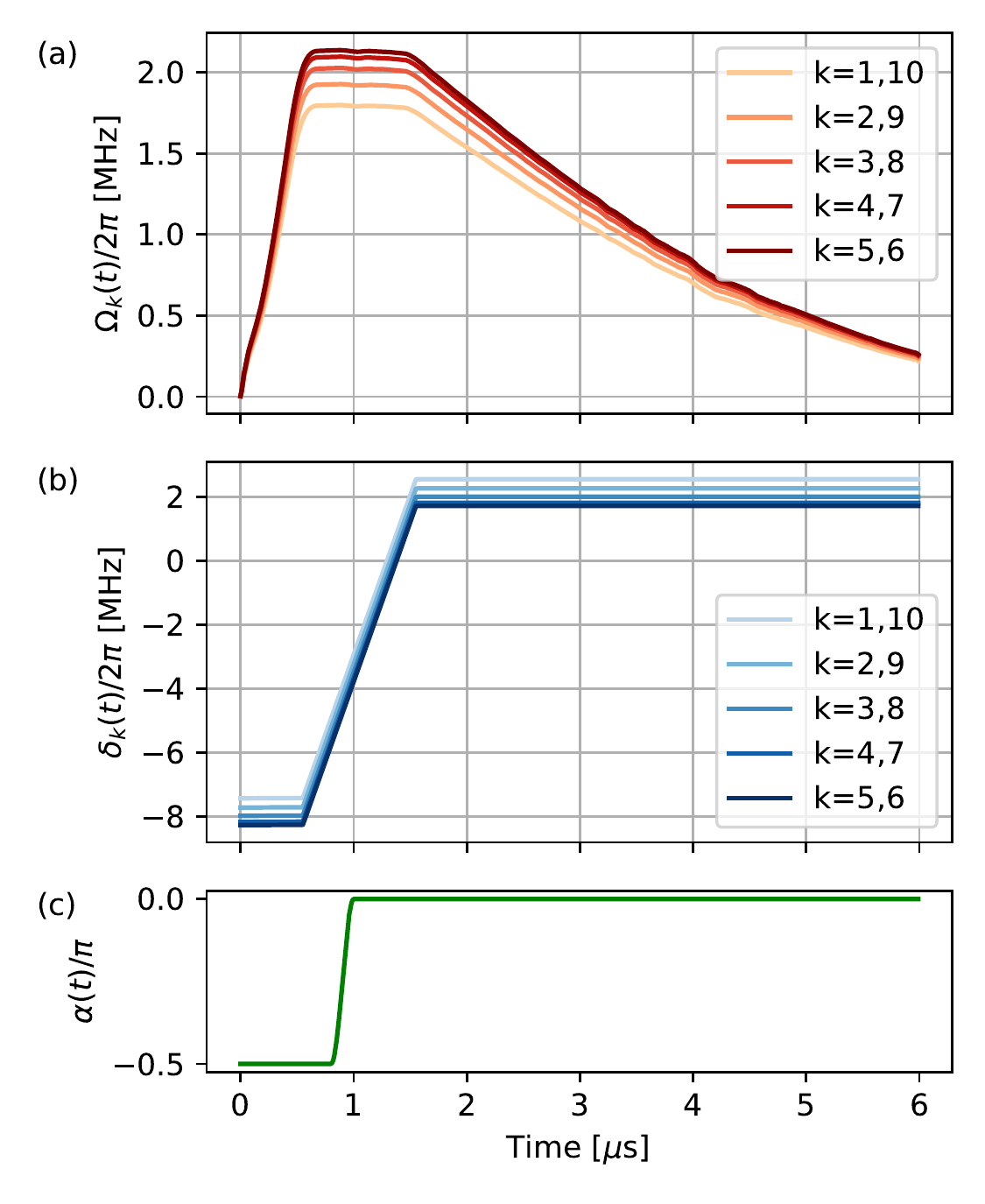}
\caption{(a,b) Control protocols of the external fields $\Omega_{k}(t)$ and $\delta_{k}(t)$. (c) Time-dependence of $\alpha(t)$, which parameterizes the time-dependent choice of the computational basis as described in the text. Initially, the quantization axis aligns with $\sigma^x$, before it is rotated on the time interval between $t_0=0.8\mu$s and $t_1=1\mu$s with $\alpha(t)=-\frac\pi2\cos^2\big(\frac\pi2(t-t_0)/(t_1-t_0)\big)$ to align with the $\sigma^z$ quantization axis.}
\label{fig:control_protocol}
\end{figure}

For the simulations we incorporate further experimental details, extending the elementary Rydberg atom Hamiltonian given in Eq.~\eqref{eq:rydberg_hamiltonian} of the main text. We include spatial laser intensity profiles that were extracted from the experimental setup such that the considered model Hamiltonian reads
\begin{align}
    H(t) = \hbar \sum_{k,l=0}^{L-1} \delta_k(t) n_{(kl)} + \frac{1}{2}\sum_{k,l=0}^{L-1} \Omega_k(t) \sigma^x_{(kl)}+ \sum_{i<j}U_{ij} n_{i}n_{j}\ .
\end{align}
Here, we introduced the notation $(kl)\equiv kL+l$ to map between double and single indices of the lattice sites; accordingly, the lasers shining in along one of the lattice dimensions exhibit an intensity profile perpendicular to that direction. The spatial and temporal form of the control fields during the considered protocol are shown in Fig.~\ref{fig:control_protocol}a,b. The coupling is $U_{ij}=U/\Delta r_{ij}^6$ with nearest neighbor interaction energy $U/h=1.947\text{MHz}$, where $h$ is Planck's constant, and $\Delta r_{(kl)(mn)}=\sqrt{(k-m)^2+(l-n)^2}$ the Euclidian distance between lattice sites.

At the beginning of the protocol all atoms are prepared in their ground state, meaning that the initial state in the spin language is a polarized state $\ket{\psi(t=0)}=\ket{\downarrow,\ldots,\downarrow}$. The initial part of the protocol mostly consists of a nearly adiabatic rotation of the polarization. This situation is difficult to address with NQS when using a fixed computational basis, because polarized states that align with the computational basis are hard to encode with NQS. Therefore, we implemented our simulation in a time-dependent frame $W(t)=\exp(-i\alpha(t)\sum_i\sigma_i^y)$ with $\alpha(t)$ as shown in Fig.~\ref{fig:control_protocol}c) such that polarizations that align with the computational basis are avoided throughout the time evolution.

\bibliography{ID.bib}
\end{document}